\documentclass[prl, twocolumn]{revtex4}%

\begin{document}

\title{Comment on ``No-Signaling Condition and Quantum Dynamics"}

\author{George Svetlichny}
 \email{svetlich@mat.puc-rio.br}
    \affiliation{Departamento de Matem\'atica, \\ Pontif{\'{\i}}cia
    Universidade Cat{\'o}lica, \\ Rio de Janeiro, RJ,  Brazil}
\maketitle

While I would concur with the main thesis of Messrs. Simon, Bu\v{z}ek, and Gisin \cite{SBG}, having previously reached and published most of the same conclusions \cite{SV},  I feel some remarks are in order. The main contention of \cite{SBG} and a part of \cite{SV}, is that  the no-signal condition (NSC) along with a set \(\cal H\) of quantum mechanical hypotheses (presumed true in standard quantum theory) imply linearity of quantum processes. For this to have any bite, the hypotheses in \(\cal H\) should somehow be more compelling and less enigmatic than linearity itself, nor should they presuppose any of the linearity one is trying to prove. 

The authors' division of phenomena into static and dynamic in is rather arbitrary. Measurement is relegated to the static, yet what can be more dynamic than the interaction of a system with an apparatus? Any realistic measurement process involves  evolution of states, and  the usual trace formula, linear in the density matrix, would be arguably irreconcilable with non-linear evolutions, if they were present in the full realistic measurement process. The assumption that the ``static" aspects of quantum mechanics have the usual description is to assume part of what one is trying to prove. Furthermore, in the coincidence experiments contemplated by the authors, when viewed from a different frame the space-like separated measuring events are not simultaneous and there is an interpolating evolution. Thus the ``static" result in one frame must be equivalent to two ``static" results and a dynamic evolution in another frame. This shows that the ``static" and dynamic aspects are intimately tied together and in the end one wonders just how much of the conclusion is already present in the hypotheses and does not need the support of NSC. In Ref.\ \cite{SV}  some of the relevant ``static" aspects are  actually {\em derived\/} from NSC and that paper's \(\cal H\).

The most critical point however is the following: for the argument in Ref.\ \cite{SBG} to procede, one must admit an objective difference between probabilistic mixtures of pure states that result in the same density matrix. Without such objectification (which cannot be local \cite{SCL}) no linearity results can be deduced. Without a clear notion of the objective nature of such differences, which should be part of \(\cal H\), it is hard to see just precisely what one is achieving. Furthermore, joining a non-local objectivity with a locality assumption such as NSC,  heightens the enigmatic nature of the assumptions.   

Recently B\'ona \cite{bo} criticized Ref. \cite{SBG} in relation to the same point of objectivity, concluding that ``\ldots the declared goals of \cite{SBG} were not attained there."  I would state that they were attained with a proper reading, but alternate readings, such as B\'ona's are also consistent.

I would be the first to admit that some of my remarks could apply to my own Ref.\ \cite{SV}. In the twelve years since I have proved my results (and the six since they were publicly available), I've had some time to ponder the lessons. What has to be included in \(\cal H\) to be able to securely conclude linearity from it and NSC is itself just as enigmatic as linearity itself. No true understanding is forthcoming from this attempt, just some hints and a greater appreciation of how remarkably are the local and the non-local joined in quantum mechanics. Because of this I've come to view  NSC as ineffective in clarifying quantum mechanics and concluded that linearity and lack of signals should {\em both\/} follow from something more basic. Ref. \cite{SCL} is an attempt:  NSC is replaced by Lorentz invariance and a form of locality, and a general quantum logic is used, but there is still a form of the projection postulate (formally present in all causal theories \cite{corc}) and an objectification (non-local) of probabilistic mixtures. It is this coexistence of local and non-local that is at the heart of the problem, and is  much more perplexing than linearity. In the end, what seems possible to me is that both linearity and causality could appear as universal emergent features of quantum gravity based on a quite general non-linear version of quantum mechanics. Ironically, the role of non-linear quantum theory would then be to finally explain the linearity of quantum mechanics as we know it.

\end{document}